\documentclass[letter]{aa}
\usepackage{graphicx}
\usepackage{txfonts}
\usepackage{float}

\titlerunning{C/O Phase Diagram and WD Core Crystallization}
\authorrunning{Blouin et al.}

\begin{document}

\title{Toward Precision Cosmochronology: A New C/O Phase Diagram for White Dwarfs}

\author{Simon Blouin\inst{1}, J\'er\^{o}me Daligault\inst{1}, Didier Saumon\inst{1}, Antoine B{\'e}dard\inst{2} and Pierre Brassard\inst{2}}

\institute{Los Alamos National Laboratory, PO Box 1663, Los Alamos, NM 87545, USA\\
              \email{sblouin@lanl.gov}
         \and
             D{\'e}partement de Physique, Universit{\'e} de Montr{\'e}al, Montr{\'e}al, QC H3C 3J7, Canada\\
             }

\abstract{
The continuous cooling of a white dwarf is punctuated by events that affect its cooling rate. Probably the most significant of those is the crystallization of its core, a phase transition that occurs once the C/O interior has cooled down below a critical temperature. This transition releases latent heat as well as gravitational energy due to the redistribution of the C and O ions during solidification, thereby slowing down the evolution of the white dwarf. The unambiguous observational signature of core crystallization---a pile-up of objects in the cooling sequence---was recently reported. However, existing evolution models struggle to quantitatively reproduce this signature, casting doubt on their accuracy when used to measure the ages of stellar populations.
The timing and amount of the energy released during crystallization depend on the exact form of the C/O phase diagram. Using the advanced Gibbs--Duhem integration method and state-of-the-art Monte Carlo simulations of the solid and liquid phases, we have obtained a very accurate version of this phase diagram, allowing a precise modeling of the phase transition. Despite this improvement, the magnitude of the crystallization pile-up remains underestimated by current evolution models. We conclude that latent heat release and O sedimentation alone are not sufficient to explain the observations and that other unaccounted physical mechanisms, possibly $^{22}$Ne phase separation, play an important role.}

\maketitle

\section{Introduction}
As a white dwarf cools down, the fully ionized plasma that makes up its core eventually becomes so correlated that a first-order phase transition occurs, leading to the formation of a solid core. Predicted more than 50 years ago \citep{vanhorn1968}, the unequivocal observational signature of this phenomenon has recently been brought to light using data from the \textit{Gaia} satellite \citep{tremblay2019}. This signature comes in the form of a pile-up of objects in the cooling sequence of evolving white dwarfs \citep[see also][]{bergeron2019}. The core crystallization of white dwarfs is accompanied by the release of latent heat and gravitational energy from the change in the C/O abundance profile \citep{mochkovitch1983,garciaberro1988,isern1997,fontaine2001,althaus2010}. Those two phenomena temporarily slow down the evolution of solidifying white dwarfs, leading to a pile-up of objects at the luminosities where the phase transition takes place.

Detecting this pile-up for normal-mass white dwarfs (\mbox{$\sim\,0.6\,M_\odot$}) is a delicate exercise as crystallization occurs at the same time as convective coupling, another important event in the evolution of a white dwarf. Convective coupling refers to the contact of the superficial convection zone with the degenerate and highly conductive interior. This temporarily slows down the evolution of the white dwarf \citep[][Figure~5]{fontaine2001}, thereby masking the delay caused by crystallization. For more massive objects however, crystallization starts well before convective coupling, which allowed \cite{tremblay2019} to unambiguously attribute to core crystallization the pile-up structure detected in the luminosity function of DA white dwarfs with masses between 0.9 and 1.1\,$M_{\odot}$.

Using white dwarf population simulations, \cite{tremblay2019} showed that the observed pile-up of massive white dwarfs is roughly consistent with the predictions of theoretical evolution sequences and that both latent heat release and C/O phase separation are needed to explain the observations.\footnote{We note that earlier results were already supporting the occurrence of O sedimentation in white dwarfs \citep{garciaberro2010}.} However, discrepancies between the models and the observations were noticed: (1) the crystallization bump is predicted to start too early and (2) the amplitude of the pile-up is significantly underestimated \citep[see also][]{kilic2020}. Such problems are a source of concerns for the use of white dwarfs as cosmochronometers \citep{winget1987,oswalt1996,fontaine2001,kalirai2012,hansen2013,tremblay2014,kilic2017,isern2019,fantin2019}. Core crystallization is a significant event in the evolution of a white dwarf and an accurate description of this phenomenon is needed to generate reliable theoretical cooling sequences.

In this letter, we present a new C/O phase diagram aimed at improving the modeling of the pile-up structure discovered by \cite{tremblay2019}. We describe our phase diagram in Section~\ref{sec:dphase} and our white dwarf population simulations in Section~\ref{sec:simul}, where we show the impact of our improved description of the phase transition on the luminosity function of massive hydrogen-atmosphere white dwarfs.

\section{C/O phase diagram}
\label{sec:dphase}
The exact shape of the C/O phase diagram is crucial to determine the impact of core crystallization on white dwarf cooling \citep[e.g.,][]{althaus2012}. The position of the liquidus dictates the temperature at the liquid--solid transition, and the separation between the liquidus and the solidus, $\Delta x_{\scriptscriptstyle \rm O}$, determines the importance of O sedimentation during crystallization. Several versions of the C/O phase diagram already exist in the literature, each with their particular limitations. Early calculations used density-functional methods \citep{segretain1993}, which are intrinsically more approximate than modern simulation techniques. Other studies relied on analytic fits to Monte Carlo (MC) simulations \citep{ogata1993,medin2010}. This approach can be delicate	 due to the use of approximate linear mixing rules and to the sensitivity of the phase diagram on the somewhat arbitrary choices that are made to construct the analytic functions used to interpolate the MC data \citep{dewitt1996}. Finally, molecular dynamics (MD) simulations were used to directly simulate the phase transition \citep{horowitz2010}. Challenges associated with those MD methods include the existence of finite-size effects and their extreme computational cost, which prohibits the detailed, well-sampled mapping of the phase diagram that is needed for white dwarf modeling.

We have computed a new, accurate phase diagram by adapting the Gibbs--Duhem integration technique coupled to MC simulations \citep{kofke1993a,kofke1993b} to plasmas. This advanced approach, previously only used for mixtures of neutral particles, was specifically designed to address the limitations of other phase diagram mapping techniques outlined above. It consists in calculating the phase diagram by integrating at constant pressure the Clapeyron equation $dT/d\xi$ along the liquid--solid coexistence curve in the temperature ($T$) -- fugacity fraction ($\xi$) space. The Clapeyron equation depends on the enthalpies and the compositions of the liquid and solid phases, which we calculate with Monte Carlo simulations in the semi-grand canonical ensemble, i.e., at constant pressure, temperature, number of ions and fugacity fraction. The details of our adaptation of the method to plasmas and analytical fits to our C/O phase diagram for an accurate implementation in white dwarf evolution codes will be published elsewhere.

\begin{figure*}
  \centering
  \includegraphics[width=1.3\columnwidth]{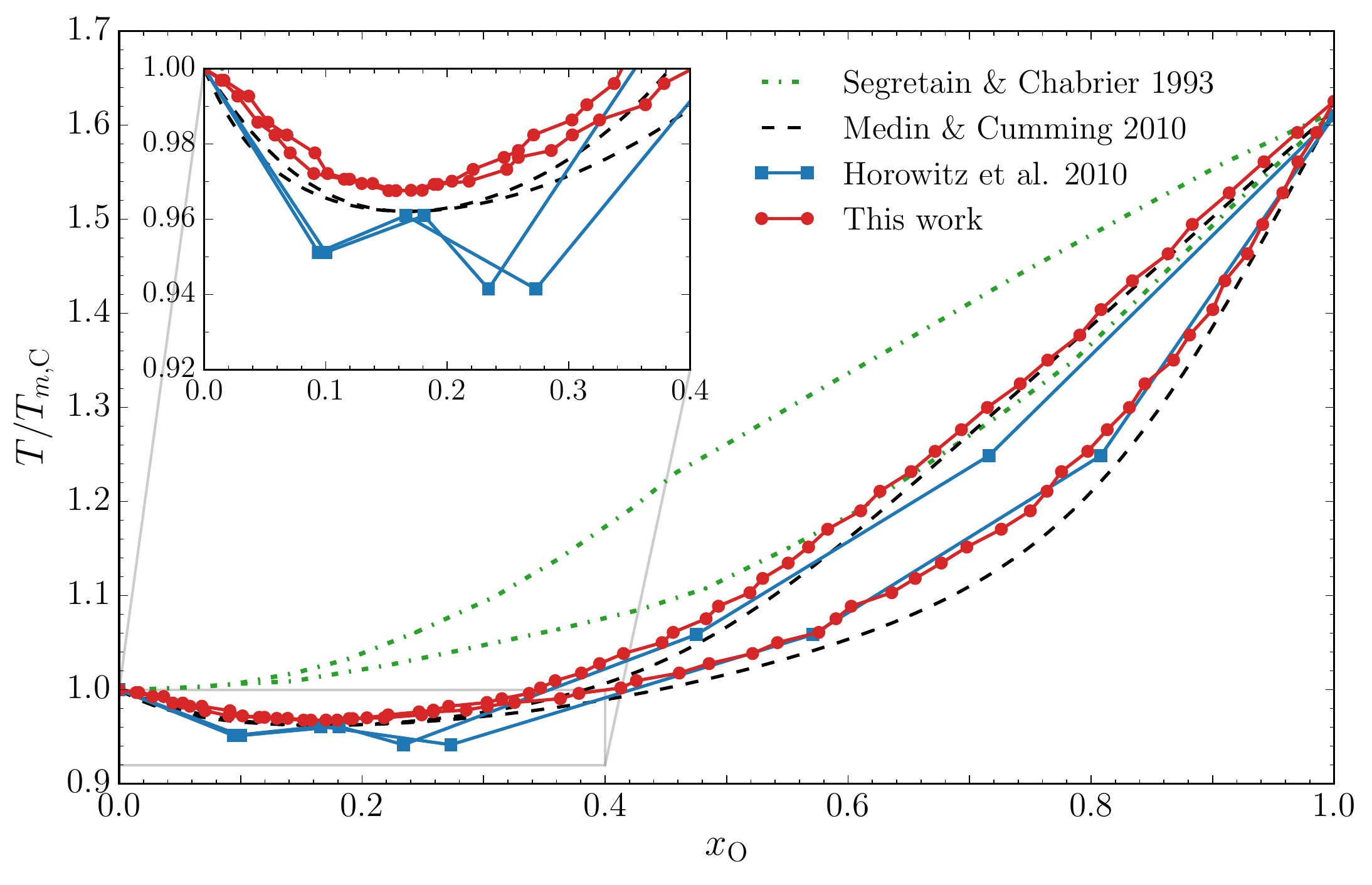}
  \caption{C/O phase diagram. Both the liquidus (above which the plasma is entirely liquid) and the solidus (below which it is entirely solid) are shown. The horizontal axis gives the number fraction of O, $x_{\scriptscriptstyle \rm O} = N_{\rm O}/ ( N_{\rm C} + N_{\rm O})$, and the vertical axis corresponds to the ratio between the temperature and the melting temperature of a pure carbon plasma. Our results, in red, are compared to those of \cite{segretain1993}, \cite{medin2010} and \cite{horowitz2010}. The region where an azeotrope is predicted is enlarged in the upper-left corner.}
  \label{fig:dphase}
\end{figure*}

Our new C/O phase diagram is shown in Figure~\ref{fig:dphase}, where we compare it to previous calculations. Qualitatively, it is close to those of \cite{medin2010} and \cite{horowitz2010}, with a similar azeotropic form. However, the quantitative differences are important in the context of white dwarf modeling. For instance, the narrow separation between the liquidus and the solidus in the phase diagram of \cite{horowitz2010} would lead to a slight underestimation of O sedimentation. We are confident that our new phase diagram is the most accurate as it is free of many of the limitations and approximations of previous studies. In particular, (1) we directly integrate the MC simulations along the coexistence curve, removing the need for arbitrary fit functions to a sparse set of simulations; (2) the relatively low cost of MC simulations allows a very fine sampling of the diagram; (3) finite-size effects are easily mitigated since there is no need to simulate a liquid--solid interface as with MD methods; (4) all calculations are performed at constant pressure and not at constant volume as in all other approaches (phase transitions occur at constant pressure); (5) the relativistic electron jellium is explicitly included in the MC simulations; (6) screening of the ion--ion interactions by relativistic electrons is accounted for; and (7) the numerical precision of the MC simulations and Gibbs--Duhem integration is revealed by the smoothness of the final coexistence curve and the recovery of the exact melting temperature at the end point of the integration at $x_{\scriptscriptstyle \rm O}=1$ (see Figure~\ref{fig:dphase}). Finally, our approach provides a full description of the thermodynamics of the phase transition.

\section{Pile-up in the cooling sequence}
\label{sec:simul}
We have computed new evolution sequences for massive hydrogen-atmosphere white dwarfs in order to test our updated phase diagram against the observed core crystallization pile-up. To do so, we used STELUM, the Montreal white dwarf evolution code \citep{brassard2018}. The constitutive physics is identical to what is described in \cite{fontaine2001}, except that (1) we now use the \cite{cassisi2007} conductive opacities, which we correct in the moderately coupled and moderately degenerate regime \citep{blouin2020} following the new theory of \cite{shaffer2020}; (2) the plasma coupling parameter at the phase transition is given by our new C/O phase diagram; (3) the release of gravitational energy due to O sedimentation is implemented following \cite{isern1997,isern2000}; (4) diffusion of $^{22}$Ne in the liquid phase is included. We assume an initially chemically homogeneous core with $X({\rm C}) = X({\rm O}) = 0.49$ and $X(^{22}{\rm Ne})=0.02$ (consistent with the results of \citealt{salaris2010} for $M_{\star} \approx 1\,M_{\odot}$) and we use an envelope stratification given by $M_{\rm H}/M_{\star}=10^{-4}$ and $M_{\rm He}/M_{\star}=10^{-2}$ (the canonical values for DA stars). We find that the cooling delay imposed by C/O phase separation is of $1.0\,$Gyr at $\log L/L_{\odot}=-4.5$ for our $0.9\,M_{\odot}$ sequence. This is close to the value obtained by \citet[Figure 5]{althaus2012} using the \cite{horowitz2010} phase diagram, which is unsurprising given the similarity between both phase diagrams (Figure~\ref{fig:dphase}).

In order to compare our cooling sequences to the luminosity function given in \cite{tremblay2019}, we have developed our own MC population synthesis code. We use the initial mass function of \cite{salpeter1955}, main-sequence lifetimes from \cite{hurley2000}, the initial--final mass relation of \cite{cummings2018} and synthetic photometry from state-of-the-art atmosphere models\footnote{\url{http://www.astro.umontreal.ca/~bergeron/CoolingModels/}} \citep{bergeron1995,kowalski2006,tremblay2011,blouin2018}. Our slightly different choice of main sequence lifetimes and initial--final mass relation from those of \cite{tremblay2019} has no effect on the results presented below.

Figure~\ref{fig:wdlf} compares our theoretical luminosity function for hydrogen-atmosphere white dwarfs between 0.9 and 1.1\,$M_{\odot}$ (black solid line) to the data of \citet[in red]{tremblay2019}. We have assumed a constant stellar formation rate and a 10\,Gyr age for the Galactic disk. The new C/O phase diagram leads to a luminosity function that is very close to that obtained by \cite{tremblay2019}. The magnitude of the crystallization pile-up is very similar and the fit to the low-luminosity cut-off is nearly identical. On this, we note that the corrections to the \cite{cassisi2007} opacities in the regime of moderate Coulomb
coupling and partial electron degeneracy \citep{blouin2020}---which significantly increase the conductivity of the H envelope---were crucial to obtain a cut-off close to the observations. \cite{tremblay2019} were able to reproduce the cut-off without those corrections, because they were relying on older calculations \citep{hubbard1969,itoh1983,mitake1984} that predict higher conductivities than \cite{cassisi2007} in the core and in the He envelope \citep{salaris2013}.

One notable difference between our luminosity function and that of \cite{tremblay2019} is that the bump associated with crystallization starts later in our simulations ($\log L/L_{\odot} \approx -2.6$ instead of $-2.3$), bringing the theoretical luminosity function closer to the data. This shift is due to the fact that the Coulomb coupling parameter at the liquid--solid transition for an equimassic C/O mixture is $\Gamma \approx 215$ according to our phase diagram, while \cite{tremblay2019} used the one-component plasma result of $\Gamma=175$ \citep{potekhin2000}. We note that this problem does not affect evolution codes that already include a detailed treatment of crystallization \citep[e.g.,][]{althaus2012}, but only a comparison with \cite{tremblay2019} is possible at this point since no other studies have yet tried to model the {\it Gaia} crystallization pile-up.

\begin{figure}
  \centering
  \includegraphics[width=\columnwidth]{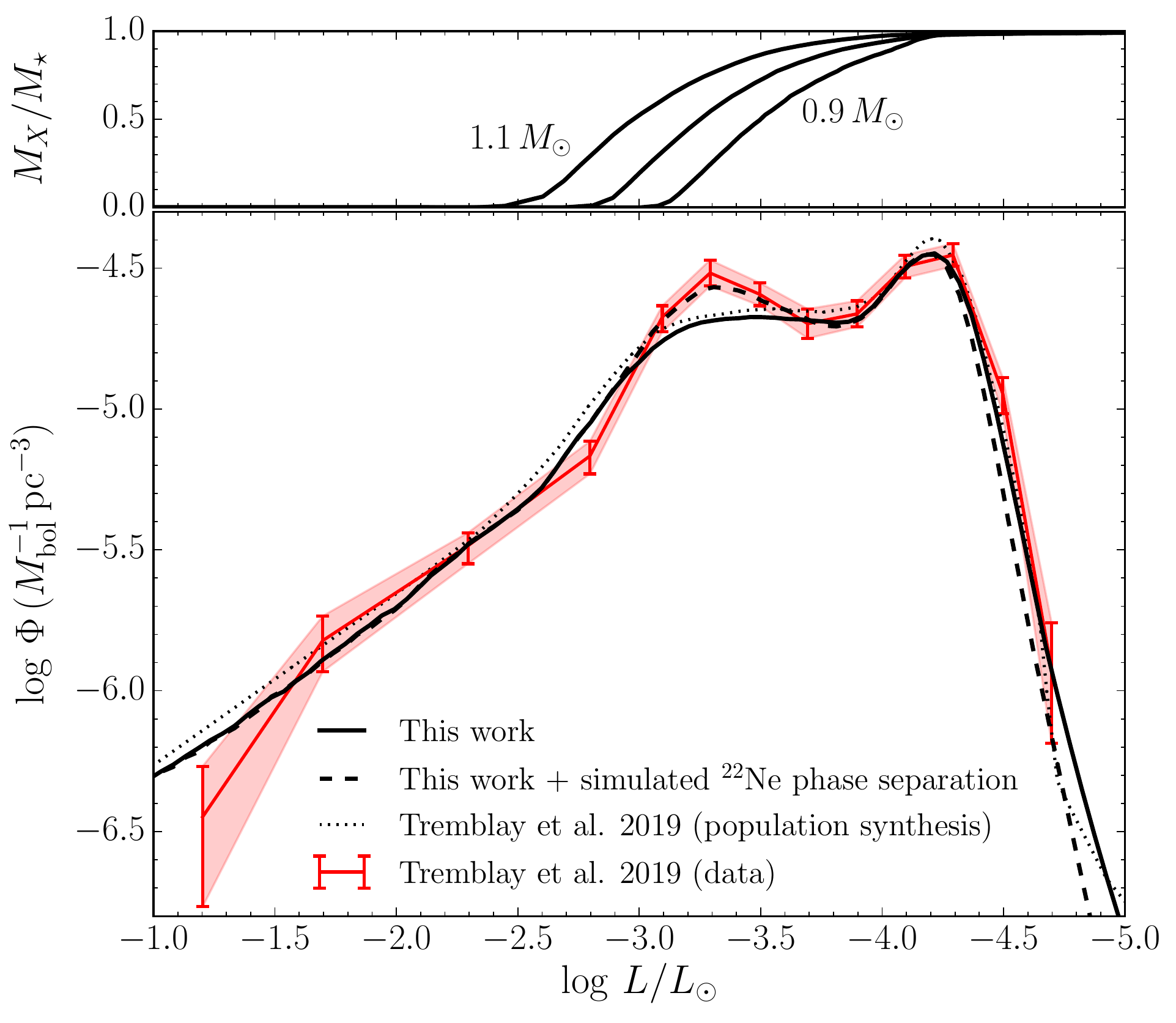}
  \caption{The lower panel displays luminosity functions for hydrogen-atmosphere white dwarfs with masses between 0.9 and 1.1\,$M_{\odot}$. Observational data from \cite{tremblay2019} are shown in red and their population synthesis is represented by the dotted curve. Results from our own simulations are shown as dashed and solid lines (with and without an attempt to simulate the effect of $^{22} {\rm Ne}$ phase separation, respectively). The normalization is arbitrary and we use the \textit{Gaia} $G$ magnitude limit given in \cite{gentile2019}. The upper panel shows the evolution of the fraction of the core that is crystallized for $1.1$, $1.0$ and $0.9\,M_{\odot}$ sequences.}
  \label{fig:wdlf}
\end{figure}

We have performed the most accurate calculation of the classical C/O phase diagram to date. Because it is in good agreement with previous calculations, we can conclude that the physics of the solidification of the C/O plasma in white dwarfs is by and large well understood. This implies that the excess in the luminosity function at $\log L/L_{\odot} \approx -3.3$ above the prediction from C/O crystallization is likely due to a separate mechanism. The gravitational settling of $^{22}$Ne in the C/O liquid phase is another source of cooling delay in white dwarfs \citep{bildsten2001,garcia-berro2008,camisassa2016}. It is already included in our sequences, but we note that its importance may be underestimated given remaining uncertainties on the initial $X(^{22}{\rm Ne})$ profile and on the diffusion coefficients \citep[e.g.,][]{cheng2019}. However, we checked that arbitrarily increasing the importance of $^{22}$Ne settling does not lead to a more prominent crystallization bump and worsens the overall agreement with the observations by reducing the number of objects that have had the time to evolve to lower luminosities. At $\log L/L_{\odot} \approx -3.3$, a significant portion of the core is already solidified (see the upper panel of Figure~\ref{fig:wdlf}), meaning that $^{22}$Ne diffusion is already largely stopped. Therefore, any change to the treatment of $^{22}$Ne settling---whether from the initial $X(^{22}{\rm Ne})$ profile or the diffusion coefficients---is unlikely to solve the discrepancy at $\log L/L_{\odot} \approx -3.3$.

Another possibly important cooling delay may arise from the phase separation of $^{22}$Ne during crystallization \citep{isern1991,althaus2010}. Our current best understanding is that at the small $^{22}$Ne concentrations typical of C/O white dwarfs ($\sim 1$\% by number), the presence of $^{22}$Ne should not affect the phase diagram, except near the azeotropic point of the C/O/Ne phase diagram. Thus the crystallization of the C/O core initially proceeds as in the case without $^{22}$Ne with no redistribution of neon ions between the solid and liquid phases. After a significant fraction of the core has crystallized, the temperature  approaches the azeotropic point and the existing calculations indicate that the liquid phase is enriched in $^{22}$Ne relative to the solid \citep{segretain1996,garcia-berro2016}. The $^{22}$Ne-poor solid is lighter than the surrounding liquid and floats upward where it eventually melts. This gradually displaces the $^{22}$Ne-rich liquid downward toward the solid--liquid interface until the azeotropic composition is reached, thereby releasing a considerable amount of gravitational energy. Given our very limited knowledge of the ternary C/O/Ne phase diagram \citep{segretain1996,hughto2012}, this effect cannot be quantitatively implemented in our evolution models. However, we note that our current understanding of $^{22}$Ne phase separation is remarkably consistent with the missing cooling delay. In Figure~\ref{fig:wdlf} we show the luminosity function obtained by adding an artificial 0.6\,Gyr delay when 60\% of the core is crystallized. Those parameters are entirely consistent with those found in preliminary studies \citep{segretain1996,garcia-berro2016} and yield an excellent fit to the crystallization pile-up.\footnote{The additional cooling delay from $^{22}$Ne phase separation worsens the fit to the low-luminosity cut-off, but this may simply be due to the unrealistic assumption that $^{22}$Ne phase separation has the same importance for all stars in our simulation. In particular, the old stars that form the cut-off of the luminosity function likely contain less $^{22}$Ne than the younger ones that form the crystallization bump, since the $^{22}$Ne abundance of a white dwarf increases for higher metallicity progenitors. Alternatively, this mismatch could be due to our assumption on the age of the Galactic disk.} Based on the current, yet limited knowledge of the C/O/Ne phase diagram, we propose that the phase separation of $^{22}$Ne in the advanced stage of crystallization significantly contributes to the pile up in the luminosity function of $0.9-1.1\,M_{\odot}$ white dwarfs (Figure~\ref{fig:wdlf}).

Finally, we speculate that the cooling anomaly for very massive white dwarfs ($1.08-1.23\,M_{\odot}$) identified by \cite{cheng2019}---where roughly 6\% of objects are affected by an unexplained $\approx 8\,{\rm Gyr}$ cooling delay---may also be at least partially explained by $^{22}$Ne phase separation rather than only by $^{22}$Ne diffusion in the liquid phase as originally suggested. In fact, the energy source responsible for the unexplained cooling delay has an effect that is highly peaked on the crystallization sequence. We found that such a peaked effect is unlikely to occur from simple diffusion alone (which is inhibited by crystallization), while it can realistically be expected from $^{22}$Ne phase separation. Of course, a 0.6\,Gyr delay is too small to explain the findings of \cite{cheng2019}, but this delay could be much more important if the initial $^{22}$Ne concentration is higher. A significant fraction of the massive objects of Cheng et al.'s sample must come from double white dwarf mergers and, interestingly, additional $^{22}$Ne is expected to be formed during merger events \citep{staff2012}. This would mean that those objects have a higher $^{22}$Ne abundance than the usual $X(^{22}{\rm Ne})=0.01-0.02$ (and a different distribution throughout the core), possibly leading to a longer cooling delay due to the phase separation of neon during crystallization.

By removing any remaining uncertainties on the classical C/O phase diagram, we have shown that the pile-up detected in the \textit{Gaia} cooling sequence cannot be explained by latent heat release and O sedimentation alone. $^{22}$Ne phase separation appears to play a crucial role in the formation of the excess of massive white dwarfs observed at $\log L/L_\odot \approx -3.3$. Our results highlight the need for a complete and accurate ternary C/O/Ne phase diagram to establish quantitatively the importance of $^{22}$Ne phase separation in white dwarf evolution. We plan to generalize our newly developed Gibbs--Duhem integration method to three-component mixtures to address this problem.

\begin{acknowledgements}
Research presented in this article was supported
by the Laboratory Directed Research and Development program of Los
Alamos National Laboratory under project number 20190624PRD2.
This work was performed under the auspices of the U.S. Department of Energy
under Contract No. 89233218CNA000001.
\end{acknowledgements}

\bibliographystyle{aa}
\bibliography{references}

\end{document}